\documentclass[acmtog]{acmart-arxiv}

\usepackage{etoolbox}
\makeatletter
\patchcmd\@combinedblfloats{\box\@outputbox}{\unvbox\@outputbox}{}{\errmessage{\noexpand patch failed}}
\makeatother

\usepackage{booktabs} 
\usepackage{import}

\usepackage[ruled]{algorithm2e} 

\SetAlFnt{\small}
\SetAlCapFnt{\small}
\SetAlCapNameFnt{\small}
\SetAlCapHSkip{0pt}

\def\fg#1{Fig.~\ref{fig:#1}}
\def\sec#1{Section~\ref{sec:#1}}

\def\permission{\relax}

\newif\ifcamera

\cameratrue

\def\revision#1{#1}

\begin{document}
\title{Interactive Video Stylization Using Few-Shot Patch-Based Training}


\def\fix#1{\goodbreak{}#1}

\ifcamera

\author{Ond\v{r}ej Texler}
\email{texleond@fel.cvut.cz}

\author{David Futschik}
\email{futscdav@fel.cvut.cz}

\author{Michal Ku\v{c}era}
\email{kucerm22@fel.cvut.cz}

\author{Ond\v{r}ej Jamri\v{s}ka}
\email{jamriond@fel.cvut.cz}

\author{\v{S}\'{a}rka Sochorov\'{a}}
\email{sochosar@fel.cvut.cz}
\affiliation{%
  \institution{\fix{Czech Technical University in Prague, Faculty of Electrical Engineering}}
  \streetaddress{Karlovo n\'{a}m\v{e}st\'{i} 13}
  \city{Praha 2}
  \state{Czech Republic}
  \postcode{121 35}
}

\author{Menglei Chai}
\email{mchai@snap.com}

\author{Sergey Tulyakov}
\email{stulyakov@snap.com}
\affiliation{%
  \institution{Snap Inc.}
  \streetaddress{2772 Donald Douglas Loop N}
  \city{Santa Monica}
  \state{United States}
  \postcode{CA 90405}
}

\author{Daniel S\'{y}kora}
\email{sykorad@fel.cvut.cz}
\affiliation{%
  \institution{Czech Technical University in Prague, Faculty of Electrical Engineering}
  \streetaddress{Karlovo n\'{a}m\v{e}st\'{i} 13}
  \city{Praha 2}
  \state{Czech Republic}
  \postcode{121 35}
}

\renewcommand\shortauthors{Texler, O.~et al.}

\else

\author{First Author}
\email{first@email.com}

\author{Second Author}
\email{second@email.com}

\author{Third Author}
\email{third@email.com}

\author{Fourth Author}
\email{fourth@email.com}

\author{Fifth Author}
\email{fifth@email.com}
\affiliation{\institution{First Institution}}

\author{Sixth Author}
\email{sixth@email.com}

\author{Seventh Author}
\email{seventh@email.com}
\affiliation{\institution{Second Institution}}

\author{Eighth Author}
\email{eighth@email.com}
\affiliation{\institution{First Institution}}

\renewcommand\shortauthors{Anonymous, A.~et al.}

\fi

\begin{abstract}

In this paper, we present a learning-based method to the keyframe-based video
stylization that allows an artist to propagate the style from a few selected
keyframes to the rest of the sequence. Its key advantage is that the resulting
stylization is semantically meaningful, i.e., specific parts of moving objects
are stylized according to the artist's intention. In contrast to previous style
transfer techniques, our approach does not require any lengthy pre-training
process nor a large training dataset. We demonstrate how to train an appearance
translation network from scratch using only a few stylized exemplars while
implicitly preserving temporal consistency. This leads to a video stylization
framework that supports real-time inference, parallel processing, and random
access to an arbitrary output frame. It can also merge the content from
multiple keyframes without the need to perform an explicit blending operation.
We demonstrate its practical utility in various interactive scenarios, where
the user paints over a selected keyframe and sees her style transferred to an
existing recorded sequence or a live video stream.


\end{abstract}

%
%


%
%


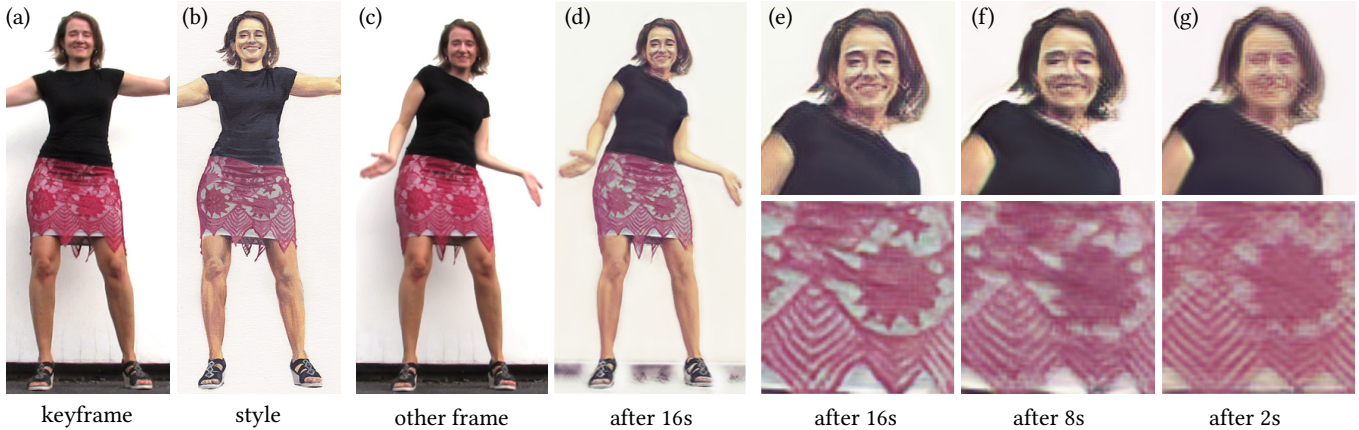
\begin{teaserfigure}
\def\svgwidth{\hsize}
\begingroup%
  \makeatletter%
  \providecommand\color[2][]{%
    \errmessage{(Inkscape) Color is used for the text in Inkscape, but the package 'color.sty' is not loaded}%
    \renewcommand\color[2][]{}%
  }%
  \providecommand\transparent[1]{%
    \errmessage{(Inkscape) Transparency is used (non-zero) for the text in Inkscape, but the package 'transparent.sty' is not loaded}%
    \renewcommand\transparent[1]{}%
  }%
  \providecommand\rotatebox[2]{#2}%
  \ifx\svgwidth\undefined%
    \setlength{\unitlength}{1427.10126953bp}%
    \ifx\svgscale\undefined%
      \relax%
    \else%
      \setlength{\unitlength}{\unitlength * \real{\svgscale}}%
    \fi%
  \else%
    \setlength{\unitlength}{\svgwidth}%
  \fi%
  \global\let\svgwidth\undefined%
  \global\let\svgscale\undefined%
  \makeatother%
  \begin{picture}(1,0.3165031)%
    \put(0,0){\includegraphics[width=\unitlength]{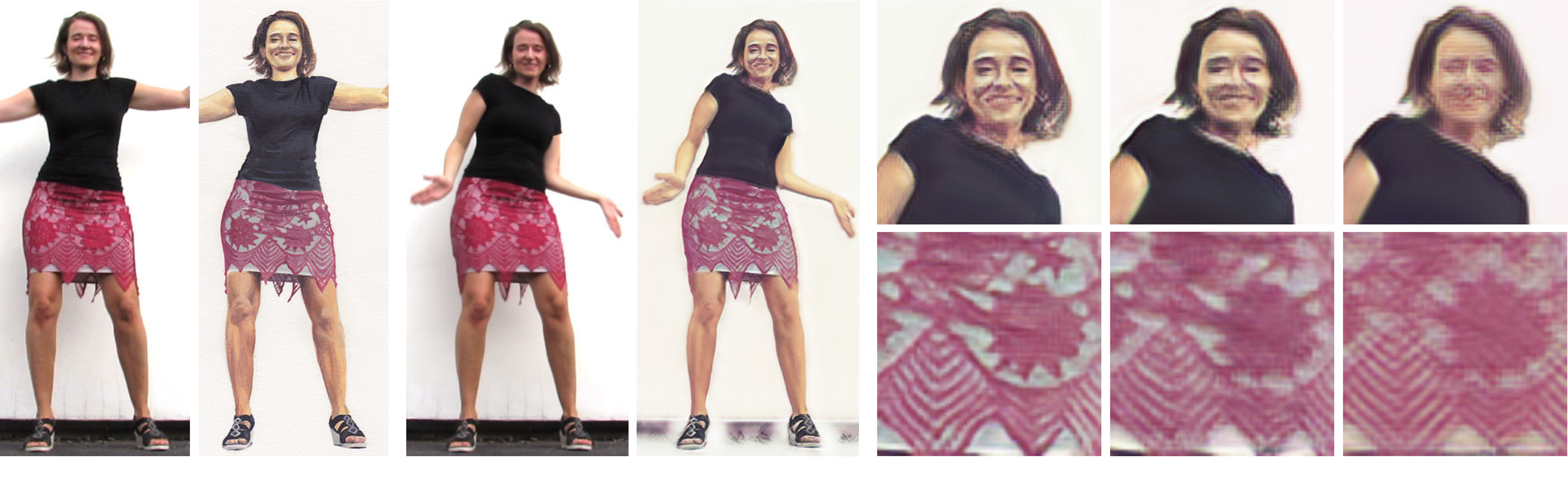}}%
    \put(0.05975569,0.00451091){\color[rgb]{0,0,0}\makebox(0,0)[b]{\smash{keyframe}}}%
    \put(0.18651272,0.00456565){\color[rgb]{0,0,0}\makebox(0,0)[b]{\smash{style}}}%
    \put(0.33002041,0.00242517){\color[rgb]{0,0,0}\makebox(0,0)[b]{\smash{other frame}}}%
    \put(0.47801271,0.00242517){\color[rgb]{0,0,0}\makebox(0,0)[b]{\smash{after 16s}}}%
    \put(0.62969953,0.00242517){\color[rgb]{0,0,0}\makebox(0,0)[b]{\smash{after 16s}}}%
    \put(0.77284016,0.00242517){\color[rgb]{0,0,0}\makebox(0,0)[b]{\smash{after 8s}}}%
    \put(0.91799783,0.00242517){\color[rgb]{0,0,0}\makebox(0,0)[b]{\smash{after 2s}}}%
    \put(0.00873882,0.3){\color[rgb]{0,0,0}\makebox(0,0)[b]{\smash{(a)}}}%
    \put(0.13991381,0.3){\color[rgb]{0,0,0}\makebox(0,0)[b]{\smash{(b)}}}%
    \put(0.26996765,0.3){\color[rgb]{0,0,0}\makebox(0,0)[b]{\smash{(c)}}}%
    \put(0.42356572,0.3){\color[rgb]{0,0,0}\makebox(0,0)[b]{\smash{(d)}}}%
    \put(0.57380033,0.3){\color[rgb]{0,0,0}\makebox(0,0)[b]{\smash{(e)}}}%
    \put(0.72403494,0.3){\color[rgb]{0,0,0}\makebox(0,0)[b]{\smash{(f)}}}%
    \put(0.87426954,0.3){\color[rgb]{0,0,0}\makebox(0,0)[b]{\smash{(g)}}}%
  \end{picture}%
\endgroup%
\caption{An example of a sequence stylized
using our approach. One frame from the original sequence is selected as a
keyframe~(a) and an artist stylizes it with acrylic paint~(b). We use this
single style exemplar as the only data to train a network. After 16 seconds of
training, the network can stylize the entire sequence in real-time~(c-d)
while maintaining the state-of-the-art visual quality and temporal coherence.
See the zoom-in views~(e-g); even after 2 seconds of training, important
structures already start to show up. Video frames~(a, c) and style exemplar~(b)
courtesy of \copyright~Zuzana Studen\'{a}\permission.}\label{fig:Zuzka1_time}
\end{teaserfigure}

\maketitle

\section{Introduction}
\label{sec:intro}

\revision{Example-based stylization of videos became recently popular thanks to
significant advances made in neural techniques~\cite{Ruder18,Sanakoyeu18,Kotovenko19b}.
Those extend the seminal approach of Gatys et al.~\shortcite{Gatys16} into the
video domain and improve the quality by adding specific style-aware content
losses.} Although these techniques can deliver impressive stylization results
on various exemplars, they still suffer from the key limitation of being
difficult to control. This is due to the fact that they only measure
statistical correlations and thus do not guarantee that specific parts of the
video will be stylized according to the artist's intention, which is an
essential requirement for use in a real production pipeline.

This important aspect is addressed by a concurrent approach---the
keyframe-based video stylization~\cite{Benard13,Jamriska19}. Those techniques
employ guided patch-based synthesis~\cite{Hertzmann01,Fiser16} to perform a
semantically meaningful transfer from a set of stylized keyframes to the rest
of the target video sequence. The great advantage of a guided scenario is that
the user has a full control over the final appearance, as she can always refine
the result by providing additional keyframes. Despite the clear benefits of
this approach, there are still some challenges that need to be resolved
to make the method suitable for a production environment.

One of the key limitations of keyframe-based stylization techniques is that
they operate in a sequential fashion, i.e., their outputs are not
\emph{seekable}. When the user seeks to any given frame, all the preceding
frames have to be processed first, before the desired result can be displayed.
This sequential processing does not fit the mechanism of how frames are handled
in professional video production tools, where random access and parallel
processing are inevitable.

Another important aspect that needs to be addressed is merging, or blending,
the stylized content from two or more (possibly inconsistent) keyframes to form
the final sequence. Although various solutions exist to this problem (e.g.,
\cite{Shechtman10,Jamriska19}), the resulting sequences usually suffer from
visible clutter or ghosting artifacts. To prevent the issues with merging, the
user has to resort to a tedious incremental workflow, where she starts by
processing the whole sequence using only a single keyframe first. Next, she
prepares a corrective keyframe by painting over the result of the previous
synthesis run. This requires re-running the synthesis after each new correction,
which leads to additional computational load and slows the overall process down.

To summarize, it would be highly beneficial to develop a guided style transfer
algorithm that would act as a fast image filter. Such a filter would perform a
semantically meaningful transfer on individual frames without the need to
access past results, while still maintaining temporal coherence. In addition,
it should also react adaptively to incoming user edits and seamlessly integrate
them on the fly without having to perform an explicit merging.

\revision{Such a setting resembles the functionality of appearance translation
networks~\cite{Isola17,Wang18}, which can give the desired look to a variety of
images and videos. In these approaches, generalization is achieved by a large
training dataset of aligned appearance exemplars. In our scenario, however, we
only have one or a few stylized examples aligned with the input video frames,
and we propagate the style to other frames with similar content. Although this
may seem like a simpler task, we demonstrate that when existing appearance
translation frameworks are applied to it naively, they lead to disturbing
visual artifacts. Those are caused by their tendency to overfit the model when
only a small set of appearance exemplars is available.}

\revision{Our scenario is also similar to few-shot learning
techniques~\cite{Liu19,Wang19} where an initial model is trained first on a
large generic dataset, and then in the inference time, additional appearance
exemplars are provided to modify the target look. Although those methods
deliver convincing results for a great variety of styles, they are limited only
to specific target domains for which large generic training datasets exist
(e.g., human bodies, faces, or street-view videos). Few-shot appearance
translation to generic videos remains an open problem.}

In this paper, we present a new appearance translation framework for arbitrary
video sequences that can deliver semantically meaningful style transfer
with temporal coherence without the need to perform any lengthy domain-specific
pre-training. We introduce a patch-based training mechanism that significantly
improves the ability of the image-to-image translation network to generalize in
a setting where larger dataset of exemplars is not available. Using our
approach, even after a couple of seconds of training, the network can
stylize the entire sequence in parallel or a live video stream in real-time.

Our method unlocks a productive workflow, where the artist provides a stylized
keyframe, and after a couple of seconds of training, she can watch the entire
video stylized. Such rapid feedback allows the user to quickly provide
localized changes and instantly see the impact on the stylized video. The
artist can even participate in an interactive session and watch how the
progress of her painting affects the target video in real-time. By replacing
the target video with a live camera feed, our method enables an unprecedented
scenario where the artist can stylize an actual live scene. When we point the
camera at the artist's face, for instance, she can simultaneously paint the
keyframe and watch a stylized video-portrait of herself. Those scenarios would
be impossible to achieve with previous keyframe-based video stylization methods,
and our framework thus opens the potential for new unconventional applications.


\section{Related Work}
\label{sec:related}

\revision{A straightforward approach to propagate the stylized content from a
painted keyframe to the rest of the sequence could be to estimate dense
correspondences between the painted keyframe and all other video
frames~\cite{Wang19c,Li19} or compute an optical flow~\cite{Chen13} between
consecutive frames, and use it to propagate the stylized content from the
keyframe. However, as shown in~Jamri\v{s}ka et al.~\shortcite{Jamriska19} this
simple approach may lead to noticeable distortion artifacts as the textural
coherence is not maintained. Moreover, even when the distortion is small the
texture advection effect leads to an unwanted perception that the stylized
content is painted on the surface.}

\revision{A more sophisticated approach to keyframe-based video stylization was
pioneered by B\'{e}nard et al.~\shortcite{Benard13} who use guided patch-based
synthesis~\cite{Hertzmann01} to maintain textural coherence. In their approach
a 3D renderer is used to produce a set of auxiliary channels, which guides the
synthesis.} This approach was recently extended to arbitrary videos by
Jamri\v{s}ka et al.~\shortcite{Jamriska19}. In their framework, guiding
channels are reconstructed automatically from the input video. Jamri\v{s}ka et
al.~also offer a post-processing step that merges the content stylized
from multiple possibly inconsistent keyframes. Although patch-based techniques
prove to deliver convincing results, their crucial drawback is that they can
stylize the video only sequentially and require an explicit merging step to be
performed when multiple keyframes are provided. Those limitations hinder random
access, parallel processing, or real-time response, which we would like to
preserve in our video stylization framework.

When considering fast video stylization, appearance translation
networks~\cite{Isola17} could provide a more appropriate solution. Once trained,
they can perform semantically meaningful appearance transfer in real-time as
recently demonstrated on human portraits~\cite{Futschik19}. Nevertheless, a
critical drawback here is that to learn such a translation network a large
training dataset is required. That can be hardly accessible in a generic video
stylization scenario, where only a few hand-drawn exemplars exist, let alone in
the context of video-to-video translation~\cite{Wang18,Chan19} which is
completely intractable.

Recently, few-shot learning techniques were introduced~\cite{Wang19b,Wang19}
to perform appearance translation without the need to have a large dataset of
specific style translation pairs. However, to do that a domain-specific dataset
is required (e.g., facial videos, human bodies in motion, etc.) to pre-train
the network. Such a requirement impedes the usage of previous few-shot methods
in a general context where the target domain is not known beforehand.

In our method, we relax the requirement of domain-specific pre-training and
show how to train the appearance translation network solely on exemplars
provided by the user. Our approach bears resemblance to previous neural texture
synthesis techniques~\cite{Li16,Ulyanov16}, which train a network with limited
receptive field on a single exemplar image and then use it to infer larger
textures that retain essential low-level characteristics of the exemplary
image. A key idea here is to leverage the fully convolutional nature of the
neural net. Even if the network is trained on a smaller patches it can be used
to synthesize larger images.

\revision{Recently, the idea of patch-based training was further explored
to accelerate training~\cite{Shocher18} or to maintain high-level
context~\cite{Zhou18,Shocher19,Shaham19}; however, all those techniques deal
only with a singe image scenario and are not directly applicable in our
context. Also, they do not use a deliberately smaller batch of randomly cropped
patches as a means of overfitting avoidance which is one of our key
contributions.}

Handling temporal consistency is a central task of video stylization methods.
When individual frames are stylized independently, the resulting stylized
animation usually contains intense temporal flickering. Although this effect is
natural for traditional hand-colored animations~\cite{Fiser14} it may become
uncomfortable for the observer when watched for a longer period of time.
\revision{Due to this reason, previous video stylization methods, either
patch-based~\cite{Benard13,Fiser17,Jamriska19,Frigo19} or
neural-based~\cite{Chen17,Sanakoyeu18,Ruder18}, try to ensure temporal
stability explicitly, e.g., by measuring the consistency between previous and a
newly generated video frame. Alternatively, blind temporal
coherency~\cite{Lai18} could be used in the post-processing step. Yet, these
approaches introduce data-dependency to the processing pipeline, which we would
like to avoid to enable random access and parallel processing.}

\revision{Our approach bears also a resemblance to a just-in-time training
recently proposed by Mullapudi et al.~\shortcite{Mullapudi19}. In their
approach, labelling is provided for a subset of frames by a more accurate
predictor and then propagated the the rest of the sequence using a quickly
trained lightweight network. To deliver sufficient quality, a relatively large
number of keyframes is necessary. Also, full-frame training is employed which
we demonstrate could suffer from strong overfitting artifacts and thus is not
applicable in our scenario where a detailed texture needs to be propagated.}




\section{Our Approach}
\label{sec:method}

The input to our method is a video sequence $I$, which consists of $N$ frames.
Optionally, every frame $I_i$ can be accompanied by a mask $M_i$ to delineate
the region of interest; otherwise, the entire video frame is stylized.
Additionally, the user also specifies a set of keyframes $I^k\subset I$, and
for each of them, the user provides stylized keyframes $S^k$, in which the
original video content is stylized. The user can stylize the entire keyframe or
only a selected subset of pixels. In the latter case, additional keyframe masks
$M^k$ are provided to determine the location of stylized regions
(see~\fg{Overview} for details).

\begin{figure}[ht]
\def\svgwidth{\hsize}\import{}{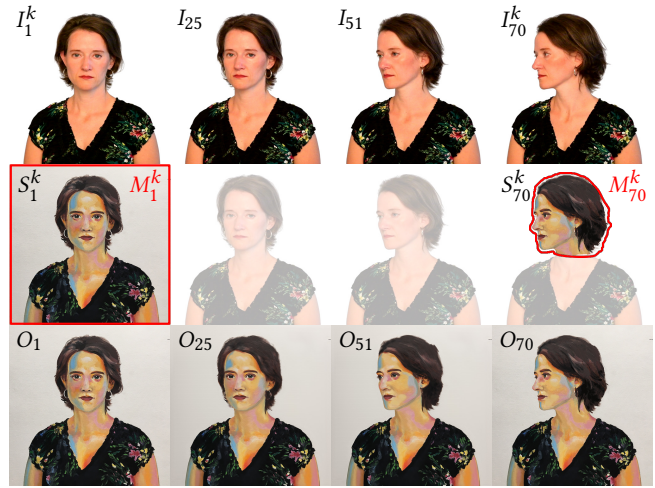}\caption{The setting of video stylization with keyframes. The
first row shows an input video sequence $I$. There are two keyframes painted by
the user, one keyframe is painted fully~($S^k_{1}$) and the other is painted
only partially~($S^k_{70}$). Mask $M^k_{1}$ denotes that the entire keyframe is
used; mask $M^k_{70}$ specifies only the head region. Our task is to stylize
all frames of the input sequence $I$ while preserving the artistic style of the
keyframes. The sequence $O$ in the bottom row shows the result of our method.
Video frames~($I$) and style exemplars~($S$) courtesy of
\copyright~Zuzana Studen\'{a}\permission.}\label{fig:Overview}
\end{figure}

\begin{figure*}[ht]
\def\svgwidth{\hsize}\import{}{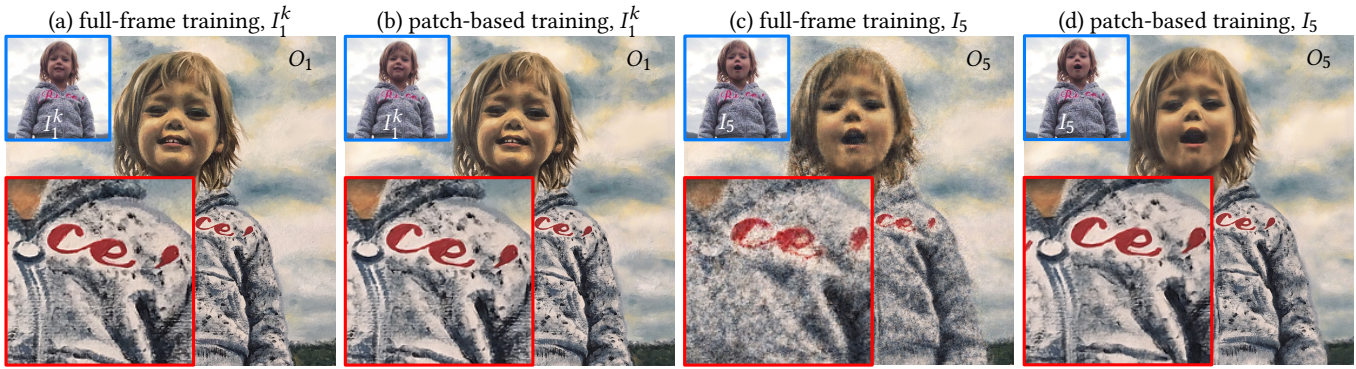}\caption{Comparison of full-frame
training vs. our patch-based approach: the original frames from the input
sequence $I$ are marked in blue and details of their stylized counterparts $O$
are marked in red. The full-frame training scheme of Futschik et
al.~\shortcite{Futschik19}~(a) as well as our patch-based approach~(b) closely
reproduce the frame on which the training was performed (see the frame
$S^k_{1}$ in~\fg{Hyper_Parameter_Optimization_scheme}). Both stylized frames~(a,
b) look nearly identical, although the training loss is lower for the
full-frame scheme. Nevertheless, the situation changes dramatically when the
two networks are used to stylize another frame from the same sequence (here
frame $I_{5}$). The network which was trained using the full-frame scheme
produces images that are very noisy and have fuzzy structure~(c). This is due
to the fact that the full-frame training causes the network to overfit the
keyframe. The network is then unable to generalize to other frames in the
sequence even though they structurally resemble the original keyframe. The
network which was trained using our patch-based scheme retains the fidelity and
preserves the important artistic details of the original style exemplar~(d).
This is thanks to the fact that our patch-based scheme better encourages the
network to generalize to unseen video frames. Video frames~($I$) courtesy of
\copyright~Zuzana Studen\'{a}\permission.}\label{fig:Full_vs_Patch_training}
\end{figure*}

\begin{figure*}[ht]
\def\svgwidth{\hsize}\import{}{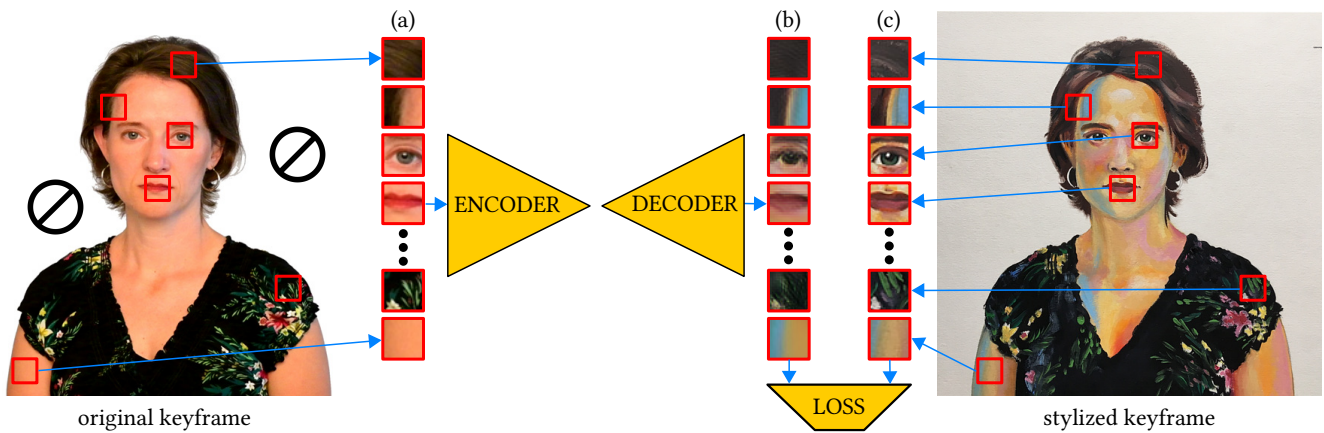}\caption{Training strategy: we randomly sample a
set of small patches from the masked area of the original keyframe~(a). These
patches are then propagated through the network in a single batch to produce
their stylized counterparts~(b). We then compute the loss of these stylized
counterparts~(b) with respect to the co-located patches sampled from the
stylized keyframe~(c) and back-propagate the error. Such a training scheme is
not limited to any particular loss function; in this paper, we use a
combination of L1 loss, adversarial loss, and VGG loss as described
in~\protect\cite{Futschik19}. Video frame~(left) and style exemplar~(right)
courtesy of \copyright~Zuzana Studen\'{a}\permission.}\label{fig:Training_scheme}
\end{figure*}

Our task is to stylize $I$ in a way that the style from $S^k$ is transferred to
the whole of $I$ in a semantically meaningful way, i.e., the stylization of
particular objects in the scene remains consistent. We denote the output
sequence by $O$. The aim is to achieve visual quality and temporal consistency
comparable to the state-of-the-art in the keyframe-based video
stylization~\cite{Jamriska19}. However, in contrast to this previous work, we
would like to stylize the video frames in random order, possibly in-parallel,
or on-demand in real-time, without the need to wait for previous frames to be
stylized or to perform explicit merging of stylized content from different
keyframes. In other words, we aim to design a translation filter that can
quickly learn the style from a few heterogeneously hand-drawn exemplars $S^k$
and then stylize the entire sequence $I$ in parallel, or any single frame on
demand. It would also be beneficial if the learning phase was fast and
incremental so that the stylization of individual video frames could start
immediately, and the stylization quality would progressively improve over time.

\begin{figure}[ht]
\def\svgwidth{\hsize}\import{}{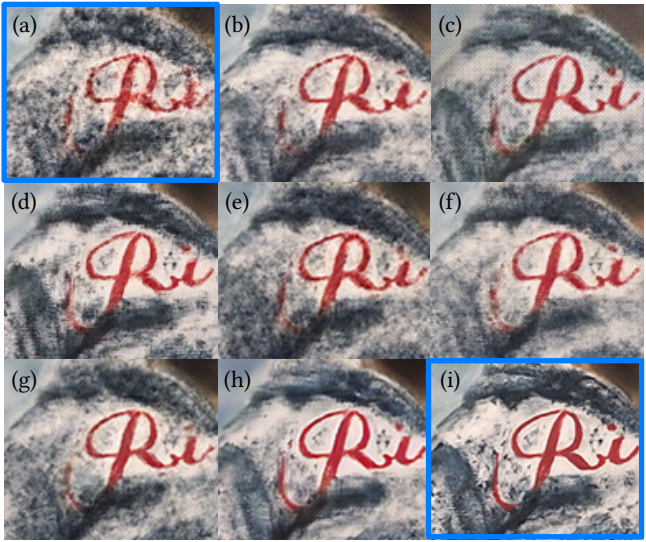}\caption{To deal with the overfitting
caused by a minimal amount of training data, we tried several commonly used
techniques to enforce regularization. In all cases shown in this figure, we
trained the network on the first frame; the shown results are zoomed details of
the fifth frame. (a)~is a result of the original full-frame training. (b-h)~are
results of full-frame training with some data augmentation. (i)~is a result of
our patch-based training strategy---see how our technique can deliver
much sharper and significantly better visual quality results, please, zoom into
the figure to better appreciate the difference. In case of~(b-c), Gaussian
noise was used to augment the data; (d)~some pixels were randomly set to black;
(e-f)~some parts of the image were occluded; (g)~dropout of entire 2D feature
maps; (h)~dropout of individual pixels before each convolution layer.}\label{fig:Other_Regularization}
\end{figure}

To design such a filter, we adopt the U-net-based image-to-image translation
framework of Futschik et al.~\shortcite{Futschik19}, which was originally
designed for the stylization of faces. It uses a custom network architecture
that can retain important high-frequency details of the original style
exemplar. Although their network can be applied in our scenario directly, the
quality of results it produces is notably inferior as compared to current
state-of-the-art (see~\fg{Full_vs_Patch_training}c and our supplementary video at 2:20).
One of the reasons why this happens is that the original Futschik et
al.'s~network is trained on a large dataset of style exemplars produced by
FaceStyle algorithm~\cite{Fiser17}. Such many exemplars are not available in
our scenario, and thus the network suffers from strong overfitting. Due to this
reason, keyframes can be perfectly reconstructed; however, the rest of the
frames are stylized poorly, even after applying well-known data augmentation
methods. See the detailed comparison in
Figures~\ref{fig:Full_vs_Patch_training} and~\ref{fig:Other_Regularization}.
Furthermore, the resulting sequence also contains a disturbing amount of
temporal flickering because the original method does not take into account
temporal coherence explicitly.

To address the drawbacks mentioned above, we alter how the network is trained
and formulate an optimization problem that allows fine-tuning the network's
architecture and its hyper-parameters to get the stylization quality comparable
to the current state-of-the-art, even with only a few training exemplars
available and within short training time. Also, we propose a solution to
suppress temporal flicker without the need to measure consistency between
individual video frames explicitly. In the following sections, those
improvements are discussed in further detail.

\subsection{Patch-Based Training Strategy}

To avoid network overfitting to the few available keyframes, we adopt a
patch-based training strategy. Instead of feeding the entire exemplar to the
network as done in~\cite{Futschik19}, we randomly sample smaller rectangular
patches from all stylized keyframes $S^k$ (see~\fg{Training_scheme}) and train
the network to predict a stylized rectangluar area of same size as input. The
sampling is performed only within the area of masked pixels $M^k$. Note that
thanks to the fully convolutional nature of the network, once trained, it can
be directly used to stylize the entire video frame even though the training was
performed on smaller patches (see~\fg{Inference_scheme}). The key benefit of
this explicit cropping and randomization step is that it simulates the scenario
when a large and diverse dataset is used for training. It prevents the network
from overfitting and generalizes to stylize the other video frames better. This
training strategy is similar to one previously used for texture
synthesis~\cite{Zhou18}.

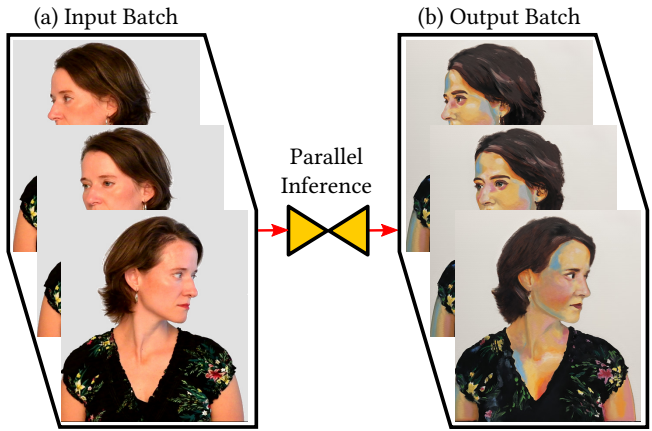
\begin{figure}[ht]
\def\svgwidth{\hsize}
\begingroup%
  \makeatletter%
  \providecommand\color[2][]{%
    \errmessage{(Inkscape) Color is used for the text in Inkscape, but the package 'color.sty' is not loaded}%
    \renewcommand\color[2][]{}%
  }%
  \providecommand\transparent[1]{%
    \errmessage{(Inkscape) Transparency is used (non-zero) for the text in Inkscape, but the package 'transparent.sty' is not loaded}%
    \renewcommand\transparent[1]{}%
  }%
  \providecommand\rotatebox[2]{#2}%
  \newcommand*\fsize{\dimexpr\f@size pt\relax}%
  \newcommand*\lineheight[1]{\fontsize{\fsize}{#1\fsize}\selectfont}%
  \ifx\svgwidth\undefined%
    \setlength{\unitlength}{790.89797974bp}%
    \ifx\svgscale\undefined%
      \relax%
    \else%
      \setlength{\unitlength}{\unitlength * \real{\svgscale}}%
    \fi%
  \else%
    \setlength{\unitlength}{\svgwidth}%
  \fi%
  \global\let\svgwidth\undefined%
  \global\let\svgscale\undefined%
  \makeatother%
  \begin{picture}(1,0.65908275)%
    \lineheight{1}%
    \setlength\tabcolsep{0pt}%
    \put(0,0){\includegraphics[width=\unitlength,page=1]{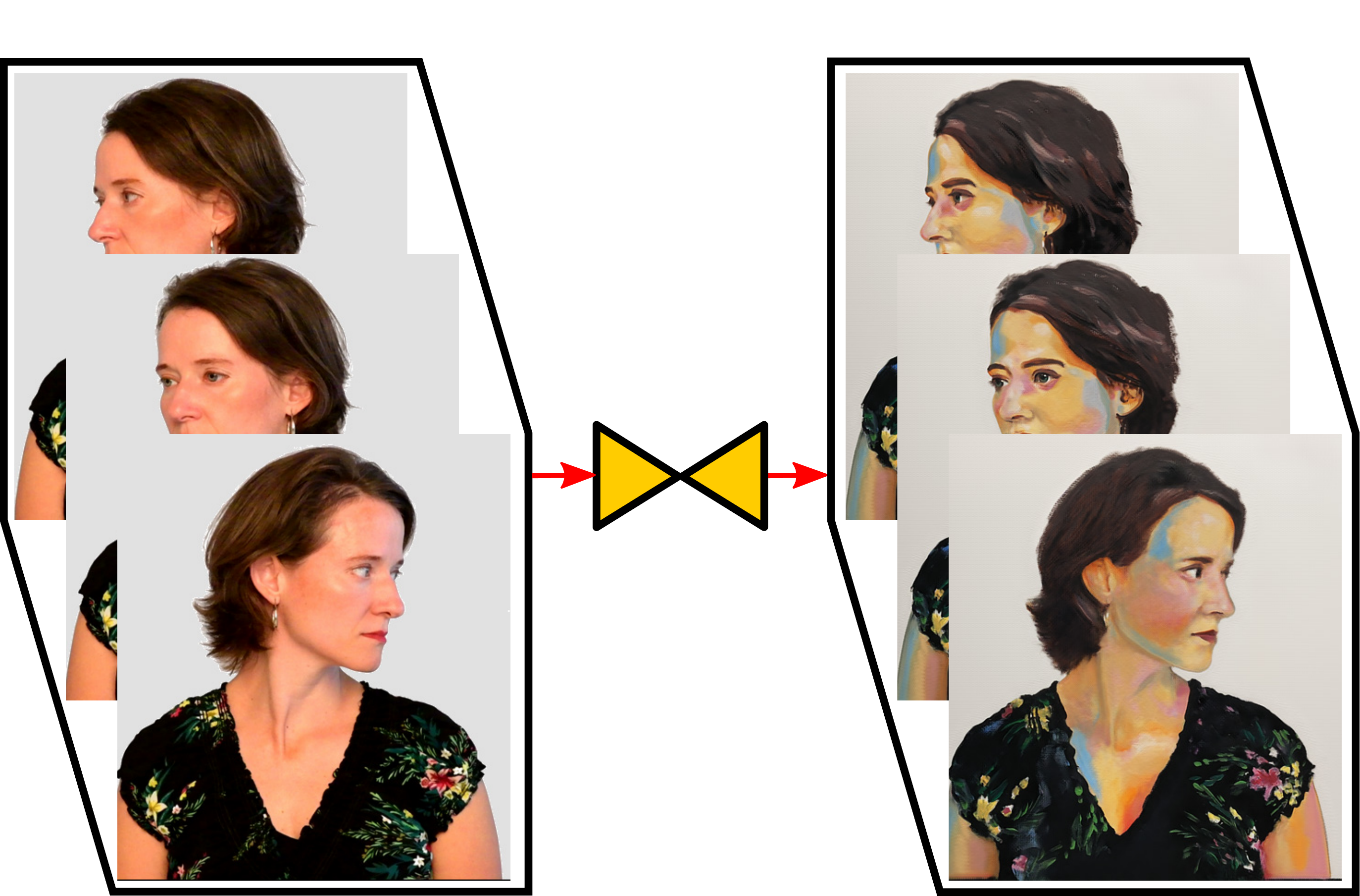}}%
    \put(0.49830792,0.4129804){\color[rgb]{0,0,0}\makebox(0,0)[t]{\lineheight{1.25}\smash{\begin{tabular}[t]{c}Parallel\\Inference\end{tabular}}}}%
    \put(0.15500455,0.63026364){\color[rgb]{0,0,0}\makebox(0,0)[t]{\lineheight{1.25}\smash{\begin{tabular}[t]{l}(a) Input Batch\end{tabular}}}}%
    \put(0.76648426,0.63026364){\color[rgb]{0,0,0}\makebox(0,0)[t]{\lineheight{1.25}\smash{\begin{tabular}[t]{l}(b) Output Batch\end{tabular}}}}%
  \end{picture}%
\endgroup%
\caption{Inference: thanks to the fully
convolutional nature of the network, we can perform the inference on entire
video frames, even though the training is done on small patches only. Since the
inference does not depend on other stylized frames, all video frames can be
stylized in parallel or in random order. This allows us to pass many or even
all of the input frames~(a) through the network in a single batch and get all
output frames~(b) at once. Video frames~(left) courtesy of
\copyright~Zuzana Studen\'{a}\permission.}\label{fig:Inference_scheme}
\end{figure}


Although the reconstruction loss measured on keyframes $S^k$ is higher when
compared to full-frame training after comparable amount of time, on the
remaining frames of $I$ the reconstruction loss is considerably lower when
comparing to the frames stylized using state-of-the-art keyframe-based video
stylization method of Jamri\v{s}ka et al.~which we purposefully consider as a
ground truth (cf.~supplementary video at 0:08 and 1:08). This lower loss w.r.t. Jamri\v{s}ka et
al. translates to much better visual quality.

\subsection{Hyper-parameter Optimization}

Although the patch-based training strategy considerably helps to resolve the
overfitting problem, we find that it is still essential to have a proper
setting of critical network hyper-parameters, as their naive values could lead
to poor inference quality, especially when the training performance is of great
importance in our applications (see~\fg{Batch_Patch_Resnet}). Besides that, we
also need to balance the model size to capture the essential characteristics of
the style yet being able to perform the inference in real-time using
off-the-shelf graphics card.

We formulate an optimization problem in which we search for an optimal setting
of the following hyper-parameters: $W_p$---size of a training patch,
$N_b$---number of patches used in one training batch, $\alpha$---learning rate,
and $N_r$---number of ResNet blocks used in our network architecture. The aim
is to minimize the loss function used in Futschik et al.~\shortcite{Futschik19}
computed over the frames inferred by our network and their counterparts
stylized using the method of Jamri\v{s}ka et al.~\shortcite{Jamriska19}. The
minimization is performed subject to the following hard constraints:
$T_t$---the time for which we allow the network to be trained for and
$T_i$---the inference time for a single video frame. Since $T_t$ as well as
$T_i$ are relatively short (in our setting $T_t=30$ and $T_i=0.06$ seconds)
full optimization of hyper-parameters becomes tractable. We used the grid
search method on a GPU cluster, to find the optimal values (see detailed
scheme~\fg{Hyper_Parameter_Optimization_scheme}). In-depth elaboration can be
found in~\sec{results}.

In our experiments, we found that hyper-parameter optimization is relatively
consistent when different validation sequences are used. We thus believe the
setting we found is useful for a greater variety of styles and sequences. Note
also that the result of Jamri\v{s}ka et al.~is used only for fine-tuning of
hyper-parameters. Once this step is finished, our framework does not require
any guided patch-based synthesis algorithm and can act fully independently.

\begin{figure}[ht]
\def\svgwidth{\hsize}\import{}{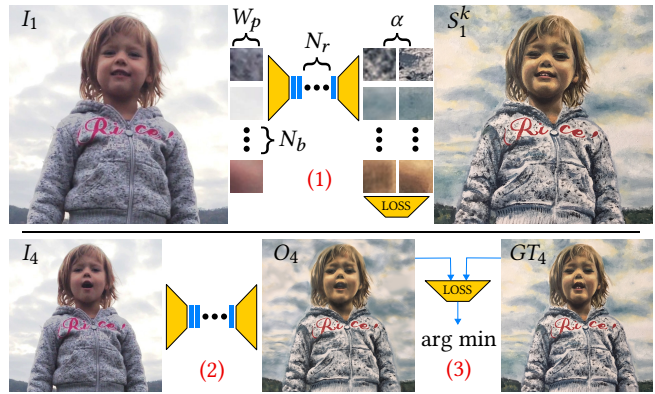}\caption{To
fine-tune critical hyper-parameters of our network, we propose the following
optimization scheme. We tune batch size $N_b$, patch size $W_p$, number of
ResNet blocks $N_r$, and learning rate $\alpha$. Using the grid search method
we sample 4-dimensional space given by these hyper-parameters and for every
hyper-parameter setting we (1)~perform a training for a given amount of time,
(2)~do inference on unseen frames, and (3)~compute the loss between inferred
frames~($O_{4}$) and result of~\cite{Jamriska19}~($GT_{4}$) - which we consider
to be ground truth. The objective is to minimize this loss. Note that the loss
in step~(1) and the loss in step~(3) are both the same. Video frames~($I$) and
style exemplar~($S$) courtesy of \copyright~Zuzana Studen\'{a}\permission.}\label{fig:Hyper_Parameter_Optimization_scheme}
\end{figure}

\subsection{Temporal Coherency}

Once the translation network with optimized hyper-parameters is trained using
the proposed patch-based scheme, style transfer to $I$ can be performed in
real-time or in parallel on the off-the-shelf graphics card. Even though such a
frame-independent process yields relatively good temporal coherence on its own
(as noted by Futschik et al.), in many cases, temporal flicker is still
apparent. We aim to suppress it while keeping the ability of the network to
perform frame-independent inference. We analyzed the source of the temporal
instability and found two main reasons: (1)~temporal noise in the original
video and (2)~visual ambiguity of the stylized content. We discuss our solution
to those issues in the following paragraphs.

We observed that the appearance translation network tends to amplify temporal
noise in the input video, i.e., even a small amount of temporal instability in
the input video causes visible flicker in the output sequence. To suppress it,
we use the motion-compensated variant of bilateral filter operating in the
temporal domain~\cite{Bennett05}. See our supplementary video (at 2:40) for the flicker
reduction that can be achieved using this pre-filtering. Although bilateral
filter requires nearby frames to be fetched into the memory, it does not
violate our requirement for frame-independent processing.

Another observation we made is that filtering the input video reduces temporal
flicker only on objects that have distinct and variable texture. Those that
lack sufficient discriminatory information (e.g., homogeneous regions) flicker
due to the fact that the visual ambiguity correlates with the network's ability
to recall the desired appearance. To suppress this phenomenon, one possibility
is to prepare the scene to contain only well distinctive regions. However, such
an adjustment may not always be feasible in practice.

Instead, we provide an additional input layer to the network that will improve
its discriminative power explicitly. This layer consists of a sparse set of
randomly distributed 2D Gaussians, each of which has a distinct randomly
generated color. Their mixture represents a unique color variation that helps
the network to identify local context and suppress the ambiguity
(see~\fg{Gnoise}). \revision{To compensate for the motion in the input video,
Gaussians are treated as points attached to a grid, which is deformed using
as-rigid-as-possible (ARAP) image registration technique~\cite{Sykora09}. In
this approach, two steps are iterated: (1)~block-matching estimates optimal
translation of each point on the grid, and (2)~rigidity is locally enforced
using the ARAP deformation model to regularize the grid structure. As this
registration scheme can be applied independently for each video frame, the
condition on frame independence is still satisfied.}

\begin{figure*}[ht]
\def\svgwidth{\hsize}\import{}{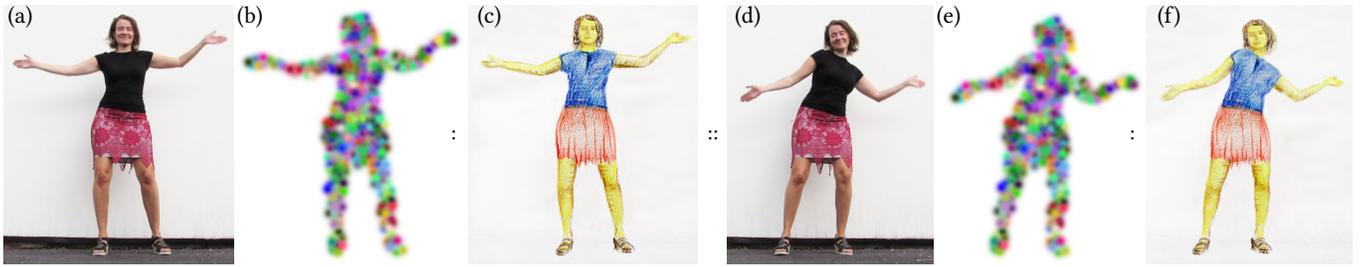}\caption{To suppress visual ambiguity of the dark mostly
homogeneous T-shirt in~(a) an auxiliary input layer is provided that contains a
mixture of randomly distributed and colored Gaussians~(b). The translation
network is trained on patches of which input pixels contain those additional
color components. The aim is to reproduce the stylized counterpart~(c). Once
the network is trained a different frame from the sequence can be stylized~(d)
using adopted version of the auxiliary input layer~(e). The resulting sequence
of stylized frames~(f) has notably better temporal stability (cf.~our
supplementary video at 2:40). Video frames~(a, d) courtesy of \copyright~Zuzana
Studen\'{a} and style exemplar~(b) courtesy of \copyright~Pavla
S\'{y}korov\'{a}\permission.}\label{fig:Gnoise}
\end{figure*}

\revision{The reason why the mixture of Gaussians is used instead of directly
encoding pixel coordinates as done, e.g., in~\cite{Liu18,Jamriska19} is the
fact that random colorization provides better localization and their sparsity,
together with rotational symmetry, reduces the effect of local distortion,
which may confuse the network. In our supplementary video (at 3:20) we, demonstrate the
benefit of using the mixture of Gaussians over the layer with color-coded pixel
coordinates. In case of extreme non-planar deformation (e.g., head rotation) or
strong occlusion (multiple scene planes), additional keyframes need to be
provided or the scene separated into multiple layers. Each keyframe or a scene
layer has then its own dedicated deformation grid. We demonstrate this scenario
in our supplementary video (at 2:56).}

\section{Results}
\label{sec:results}

We implemented our approach in C++ and Python with PyTorch, adopting the
structure of the appearance translation network of Futschik et
al.~\shortcite{Futschik19} and used their recommended settings including
training loss. Ground truth stylized sequences for hyper-parameter tuning and
comparison were produced using the video stylization method of Jamri\v{s}ka et
al.~\shortcite{Jamriska19}.

\begin{figure*}[ht]
\def\svgwidth{\hsize}\import{}{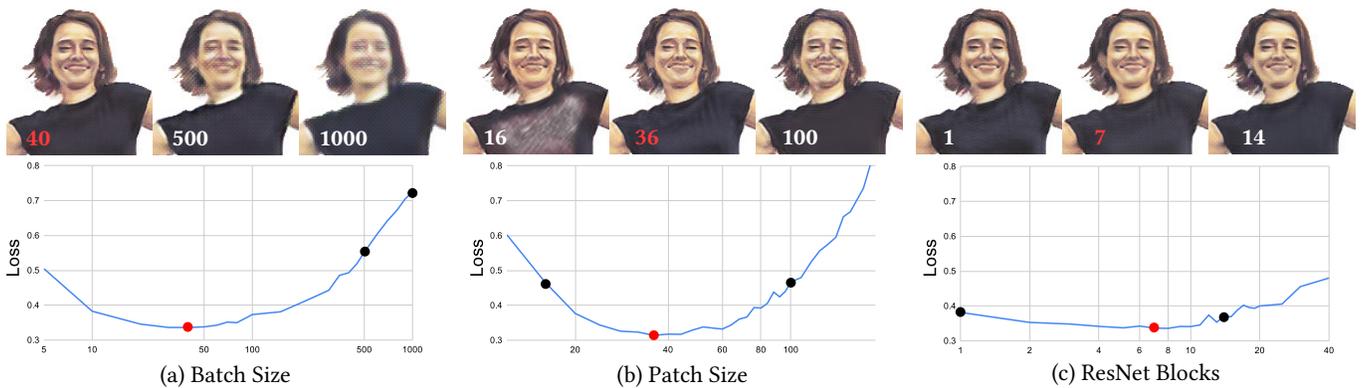}\caption{Influence of important
hyper-parameters on visual quality of results. The loss, y-axes, is computed
w.r.t.~the output of~Jamri\v{s}ka et al.~\shortcite{Jamriska19}. The best
setting for each hyper-parameter is highlighted in red: (a)~The loss curve for
the batch size $N_b$---the number of patches in one training batch (other
hyper-parameters are fixed). As can be seen, increasing $N_b$ deteriorates
visual quality significantly; it indicates that there exists an ideal amount of
data to pass through the network during the back-propagation step. (b)~The loss
curve for the patch size~$W_p$. The optimal size of a patch is around 36x36
pixels. This fact indicates that smaller patches may not provide sufficient
context while larger ones could make the network less robust to deformation
changes. (c)~The loss curve for the number of ResNet blocks $N_r$ that
corresponds to the capacity of the network. As can be seen, settings with 7
ResNet blocks is slightly better than other results; however, this
hyper-parameter does have major impact on the quality of results. \revision{For
additional experiments with hyper-parameter setting, refer to our supplementary
text.}}\label{fig:Batch_Patch_Resnet}
\end{figure*}

We performed fine-tuning of hyper-parameters on a selection of frames from our
evaluation sequences. We computed their stylized counterparts using the method
of Jamri\v{s}ka et al.~\shortcite{Jamriska19} and performed optimization using
grid search on a cluster with 48 Nvidia Tesla V100 GPUs in 3 days. We searched
over the following intervals: $W_p\in(12,188)$, $N_b\in(5,1000)$, $N_r\in(1,40)
$, $\alpha\in(0.0002,0.0032)$. In total we sampled around 200,000 different
settings of those hyper-parameters. We found the optimal patch size to
be~$W_p=36$ pixels, the number of patches in one batch~$N_b=40$, learning
rate~$\alpha=0.0004$, and the number of ResNet blocks~$N_r=7$.

See~\fg{Batch_Patch_Resnet} to compare visual quality for different
hyper-parameter settings. Note the substantial improvement in visual quality
over different settings, which confirms the necessity of this optimization. An
interesting outcome of the proposed hyper-parameter optimization is a
relatively small number of patches in one batch~$N_b=40$
(\fg{Batch_Patch_Resnet}a). This value interplays with our choice of
patch-based training scheme. Although a common strategy would be to
enlarge~$N_b$ as much as possible to utilize GPU capability, in our case,
increasing~$N_b$ is actually counterproductive as it turns training scheme into
a full-frame scenario that tends to overfit the network on the keyframe and
produce poor results on unseen video frames. A smaller number of randomly
selected patches in every batch increases the variety of back-propagation
gradients and thus encourages the network to generalize better. From the
optimal patch size~$W_p=36$ (\fg{Batch_Patch_Resnet}b) it is apparent that
smaller patches may not provide sufficient context, while larger patches may
make the network less resistant to appearance changes caused by deformation of
the target object and less sensitive to details. Surprisingly, the number of
ResNet blocks~$N_r$ (see~\fg{Batch_Patch_Resnet}c) does not have a significant
impact on the quality, although there is a subtle saddle point visible. Similar
behavior also holds true for the learning rate parameter~$\alpha$. \revision{In
addition, we also examined the influence of the number of network filters on
the final visual quality (see our supplementary material). The measurements
confirmed that the number of filters needs to be balanced as well to capture
the stylized content while still avoid overfitting.}

With all optimized hyper-parameters, a video sequence of resolution $640\times
640$ with 10\% of active pixels (inside the mask $M^k$) can be stylized in good
quality at 17 frames per second after 16 seconds of training (see~\fg{Zuzka1_time}).

\begin{figure}[!ht]
\def\svgwidth{\hsize}
\begingroup%
  \makeatletter%
  \providecommand\color[2][]{%
    \errmessage{(Inkscape) Color is used for the text in Inkscape, but the package 'color.sty' is not loaded}%
    \renewcommand\color[2][]{}%
  }%
  \providecommand\transparent[1]{%
    \errmessage{(Inkscape) Transparency is used (non-zero) for the text in Inkscape, but the package 'transparent.sty' is not loaded}%
    \renewcommand\transparent[1]{}%
  }%
  \providecommand\rotatebox[2]{#2}%
  \newcommand*\fsize{\dimexpr\f@size pt\relax}%
  \newcommand*\lineheight[1]{\fontsize{\fsize}{#1\fsize}\selectfont}%
  \ifx\svgwidth\undefined%
    \setlength{\unitlength}{960bp}%
    \ifx\svgscale\undefined%
      \relax%
    \else%
      \setlength{\unitlength}{\unitlength * \real{\svgscale}}%
    \fi%
  \else%
    \setlength{\unitlength}{\svgwidth}%
  \fi%
  \global\let\svgwidth\undefined%
  \global\let\svgscale\undefined%
  \makeatother%
  \begin{picture}(1,0.42031255)%
    \lineheight{1}%
    \setlength\tabcolsep{0pt}%
    \put(0,0){\includegraphics[width=\unitlength,page=1]{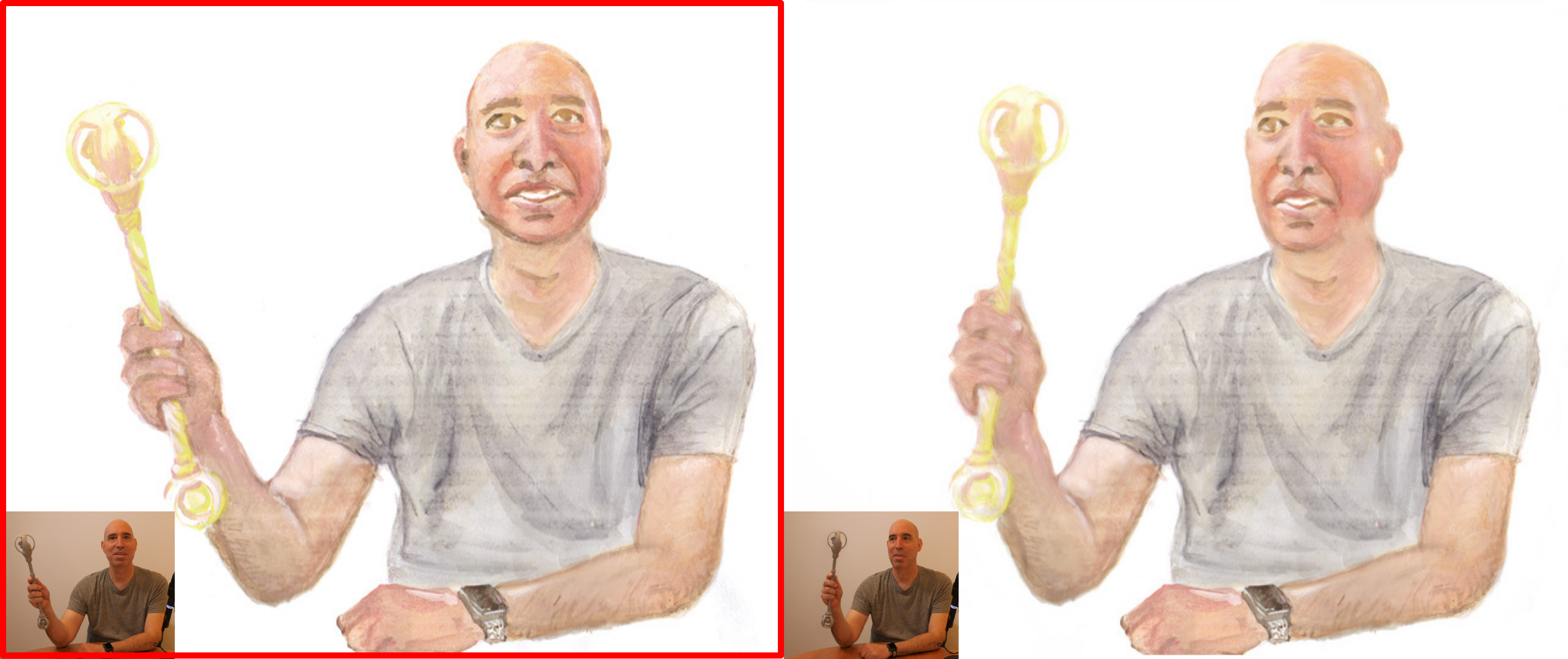}}%
    \put(0.03387454,0.37631838){\color[rgb]{0,0,0}\makebox(0,0)[t]{\lineheight{0}\smash{\begin{tabular}[t]{c}(a)\end{tabular}}}}%
    \put(0,0){\includegraphics[width=\unitlength,page=2]{Res_single_Adam.pdf}}%
    \put(0.53287357,0.37631838){\color[rgb]{0,0,0}\makebox(0,0)[t]{\lineheight{0}\smash{\begin{tabular}[t]{c}(b)\end{tabular}}}}%
    \put(0,0){\includegraphics[width=\unitlength,page=3]{Res_single_Adam.pdf}}%
  \end{picture}%
\endgroup%
\caption{\revision{Given one keyframe~(a) and a
video sequence~(in blue), our method produces the stylized result~(b). Video
frames~(insets of a, b) courtesy of \copyright~Adam Finkelstein and style
exemplars~(a) courtesy of \copyright~Pavla S\'{y}korov\'{a}\permission.}}\label{fig:Res_single_Adam}
\end{figure}

\begin{figure*}[ht]
\def\svgwidth{\hsize}
\begingroup%
  \makeatletter%
  \providecommand\color[2][]{%
    \errmessage{(Inkscape) Color is used for the text in Inkscape, but the package 'color.sty' is not loaded}%
    \renewcommand\color[2][]{}%
  }%
  \providecommand\transparent[1]{%
    \errmessage{(Inkscape) Transparency is used (non-zero) for the text in Inkscape, but the package 'transparent.sty' is not loaded}%
    \renewcommand\transparent[1]{}%
  }%
  \providecommand\rotatebox[2]{#2}%
  \newcommand*\fsize{\dimexpr\f@size pt\relax}%
  \newcommand*\lineheight[1]{\fontsize{\fsize}{#1\fsize}\selectfont}%
  \ifx\svgwidth\undefined%
    \setlength{\unitlength}{2302.10547259bp}%
    \ifx\svgscale\undefined%
      \relax%
    \else%
      \setlength{\unitlength}{\unitlength * \real{\svgscale}}%
    \fi%
  \else%
    \setlength{\unitlength}{\svgwidth}%
  \fi%
  \global\let\svgwidth\undefined%
  \global\let\svgscale\undefined%
  \makeatother%
  \begin{picture}(1,0.17592593)%
    \lineheight{1}%
    \setlength\tabcolsep{0pt}%
    \put(0,0){\includegraphics[width=\unitlength,page=1]{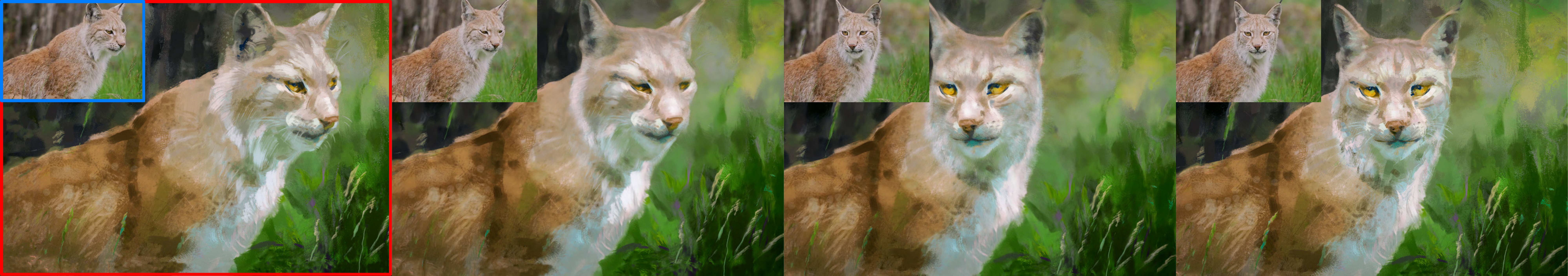}}%
    \put(0.23446906,0.01056783){\color[rgb]{1,1,1}\makebox(0,0)[t]{\lineheight{0}\smash{\begin{tabular}[t]{c}(a)\end{tabular}}}}%
    \put(0.48446904,0.01056783){\color[rgb]{1,1,1}\makebox(0,0)[t]{\lineheight{0}\smash{\begin{tabular}[t]{c}(b)\end{tabular}}}}%
    \put(0.73562968,0.01056783){\color[rgb]{1,1,1}\makebox(0,0)[t]{\lineheight{0}\smash{\begin{tabular}[t]{c}(c)\end{tabular}}}}%
    \put(0.98441811,0.01056783){\color[rgb]{1,1,1}\makebox(0,0)[t]{\lineheight{0}\smash{\begin{tabular}[t]{c}(d)\end{tabular}}}}%
    \put(0,0){\includegraphics[width=\unitlength,page=2]{Res_multi_Lynx.pdf}}%
  \end{picture}%
\endgroup%
\caption{The Lynx sequence stylized using two
keyframes~(a, d). Notice how our method produces seamless transition between
the keyframes while preserving fine texture of the style~(b, c). Watch our
supplementary video (at 1:22) to see the sequence in motion. Style exemplars~(a, d)
courtesy of \copyright~Jakub Javora\permission.}\label{fig:Res_multi_Lynx}
\end{figure*}

\begin{figure*}[ht]
\def\svgwidth{\hsize}
\begingroup%
  \makeatletter%
  \providecommand\color[2][]{%
    \errmessage{(Inkscape) Color is used for the text in Inkscape, but the package 'color.sty' is not loaded}%
    \renewcommand\color[2][]{}%
  }%
  \providecommand\transparent[1]{%
    \errmessage{(Inkscape) Transparency is used (non-zero) for the text in Inkscape, but the package 'transparent.sty' is not loaded}%
    \renewcommand\transparent[1]{}%
  }%
  \providecommand\rotatebox[2]{#2}%
  \newcommand*\fsize{\dimexpr\f@size pt\relax}%
  \newcommand*\lineheight[1]{\fontsize{\fsize}{#1\fsize}\selectfont}%
  \ifx\svgwidth\undefined%
    \setlength{\unitlength}{1919.99994233bp}%
    \ifx\svgscale\undefined%
      \relax%
    \else%
      \setlength{\unitlength}{\unitlength * \real{\svgscale}}%
    \fi%
  \else%
    \setlength{\unitlength}{\svgwidth}%
  \fi%
  \global\let\svgwidth\undefined%
  \global\let\svgscale\undefined%
  \makeatother%
  \begin{picture}(1,0.28906281)%
    \lineheight{1}%
    \setlength\tabcolsep{0pt}%
    \put(0,0){\includegraphics[width=\unitlength,page=1]{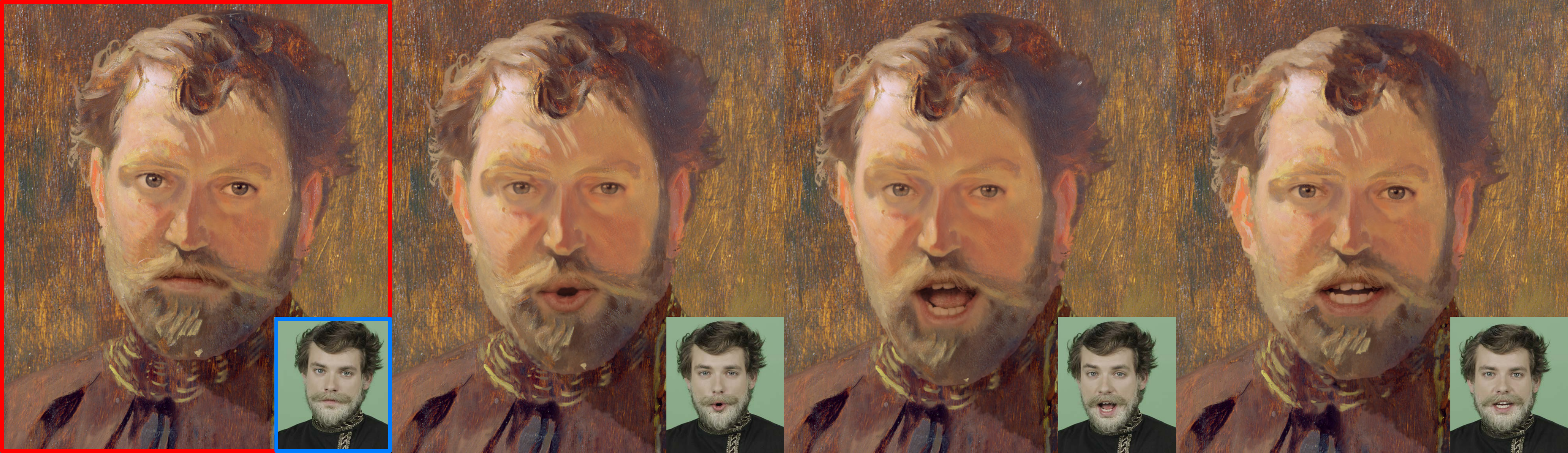}}%
    \put(0.017515,0.265735){\color[rgb]{1,1,1}\makebox(0,0)[t]{\lineheight{0}\smash{\begin{tabular}[t]{c}(a)\end{tabular}}}}%
    \put(0.26764601,0.265735){\color[rgb]{1,1,1}\makebox(0,0)[t]{\lineheight{0}\smash{\begin{tabular}[t]{c}(b)\end{tabular}}}}%
    \put(0.51702671,0.265735){\color[rgb]{1,1,1}\makebox(0,0)[t]{\lineheight{0}\smash{\begin{tabular}[t]{c}(c)\end{tabular}}}}%
    \put(0.76764602,0.265735){\color[rgb]{1,1,1}\makebox(0,0)[t]{\lineheight{0}\smash{\begin{tabular}[t]{c}(d)\end{tabular}}}}%
    \put(0,0){\includegraphics[width=\unitlength,page=2]{Res_multi_Mucha.pdf}}%
  \end{picture}%
\endgroup%
\caption{ \revision{An example sequence of 228
video frames~(in blue) as stylized from two keyframes~(a, d). Results of our
method~(b, c) stay true to style exemplars over the course of the sequence.
Video frames~(insets of a--d) and style exemplars~(a, d) courtesy of
\copyright~Muchalogy\permission.}}\label{fig:Res_multi_Mucha}
\end{figure*}

We evaluated our approach on a set of video sequences with different
resolutions ranging from $350\times 350$ to $960\times 540$, containing
different visual content (faces, human bodies, animals), and various artistic
styles (oil paint, acrylic paint, chalk, color pencil, markers, and digital
image). Simpler sequences were stylized using only one keyframe (see
Figures~\ref{fig:Zuzka1_time}, \ref{fig:Full_vs_Patch_training},
\ref{fig:Gnoise}, \ref{fig:Res_single_Adam}, and~\ref{fig:Res_single_Joli})
while the more complex ones have multiple (ranging from two to seven, see
Figures~\ref{fig:Res_multi_Imperial}, \ref{fig:Res_multi_Lynx},
\ref{fig:Res_multi_Krobot}, and~\ref{fig:Res_multi_Mucha}). Before training,
the target sequence was pre-filtered using the bilateral temporal filter. In
case that the sequence contains regions having ambiguous appearances, we
compute an auxiliary input layer with the mixture of randomly colored Gaussians
that follows the motion in the target sequence. During the training phase, we
randomly sample patches inside the mask $M^k$ from all keyframes $k$ and feed
them in batches to the network to compute the loss and backpropagate the error.
Training, as well as inference, were performed on Nvidia RTX 2080 GPU. The
training time was set to be proportional to the number of input patches (number
of pixels inside the mask $M^k$), e.g., 5 minutes for a $512\times 512$
keyframe with all pixels inside the mask. After training, the entire sequence
can be stylized at the speed of roughly 17 frames per second. See our
supplementary video (at 0:08 and 1:08) for the resulting stylized sequences.

\subsection{Comparison}

To confirm the importance of our patch-based training strategy, we conducted
comparisons with other commonly used methods for data-augmentation that can
help avoiding overfitting such as adding Gaussian noise to the input, randomly
erasing selected pixels, occluding larger parts of the input image, or
performing dropout before each convolution layer. We found that none of these
techniques can achieve comparable visual quality to our patch-based training
strategy (see~\fg{Other_Regularization}).

\begin{figure}[!ht]
\def\svgwidth{\hsize}
\begingroup%
  \makeatletter%
  \providecommand\color[2][]{%
    \errmessage{(Inkscape) Color is used for the text in Inkscape, but the package 'color.sty' is not loaded}%
    \renewcommand\color[2][]{}%
  }%
  \providecommand\transparent[1]{%
    \errmessage{(Inkscape) Transparency is used (non-zero) for the text in Inkscape, but the package 'transparent.sty' is not loaded}%
    \renewcommand\transparent[1]{}%
  }%
  \providecommand\rotatebox[2]{#2}%
  \newcommand*\fsize{\dimexpr\f@size pt\relax}%
  \newcommand*\lineheight[1]{\fontsize{\fsize}{#1\fsize}\selectfont}%
  \ifx\svgwidth\undefined%
    \setlength{\unitlength}{1012.50008651bp}%
    \ifx\svgscale\undefined%
      \relax%
    \else%
      \setlength{\unitlength}{\unitlength * \real{\svgscale}}%
    \fi%
  \else%
    \setlength{\unitlength}{\svgwidth}%
  \fi%
  \global\let\svgwidth\undefined%
  \global\let\svgscale\undefined%
  \makeatother%
  \begin{picture}(1,0.36296309)%
    \lineheight{1}%
    \setlength\tabcolsep{0pt}%
    \put(0,0){\includegraphics[width=\unitlength,page=1]{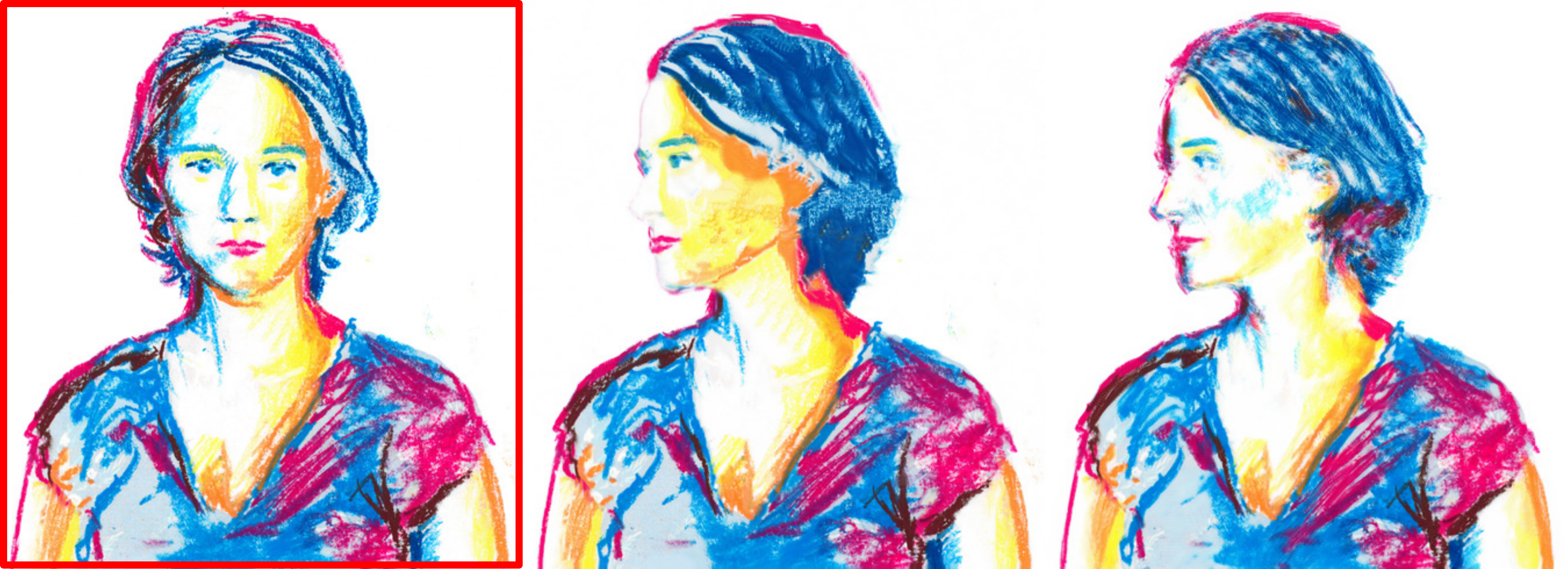}}%
    \put(0.03291557,0.32168933){\color[rgb]{0,0,0}\makebox(0,0)[t]{\lineheight{0}\smash{\begin{tabular}[t]{c}(a)\end{tabular}}}}%
    \put(0.36532293,0.32168933){\color[rgb]{0,0,0}\makebox(0,0)[t]{\lineheight{0}\smash{\begin{tabular}[t]{c}(b)\end{tabular}}}}%
    \put(0,0){\includegraphics[width=\unitlength,page=2]{Lim_Zuzka2_head_rotate.pdf}}%
    \put(0.699537,0.32168933){\color[rgb]{0,0,0}\makebox(0,0)[t]{\lineheight{0}\smash{\begin{tabular}[t]{c}(c)\end{tabular}}}}%
    \put(0,0){\includegraphics[width=\unitlength,page=3]{Lim_Zuzka2_head_rotate.pdf}}%
    \put(-0.16813427,0.4399775){\color[rgb]{0,0,0}\makebox(0,0)[lt]{\begin{minipage}{0.81029192\unitlength}\raggedright \end{minipage}}}%
  \end{picture}%
\endgroup%
\caption{When the target subject
undergoes a substantial appearance change, the results of both Jamri\v{s}ka et
al.~\shortcite{Jamriska19}~(b) and our method~(c) exhibit noticeable artifacts.
The parts that were not present in the keyframe are reconstructed poorly---see
the face and hair regions where~\cite{Jamriska19} produces large flat areas,
while our approach does not reproduce the color of the face well. Video
frames~(insets of a--c) and style exemplars~(a) courtesy of
\copyright~Zuzana Studen\'{a}\permission.}\label{fig:Lim_Zuzka2_head_rotate}
\end{figure}

\begin{figure}[!ht]
\def\svgwidth{\hsize}\import{}{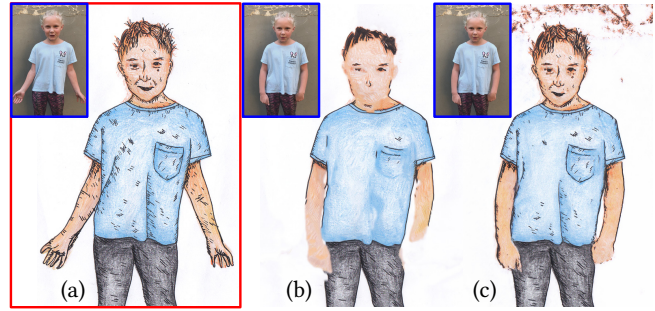}\caption{For the state-of-the-art algorithm
of~\cite{Jamriska19}, contour based styles~(a) present a particular
challenge~(b). Using our approach~(c), the contours are transferred with finer
detail and remain sharp even as the sequence undergoes transformations. Video
frames~(insets of a--c) and style exemplar~(a) courtesy of
\copyright~\v{S}t\v{e}p\'{a}nka S\'{y}korov\'{a}\permission.}\label{fig:Res_single_Joli}
\end{figure}

\begin{figure*}[ht]
\def\svgwidth{\hsize}\import{}{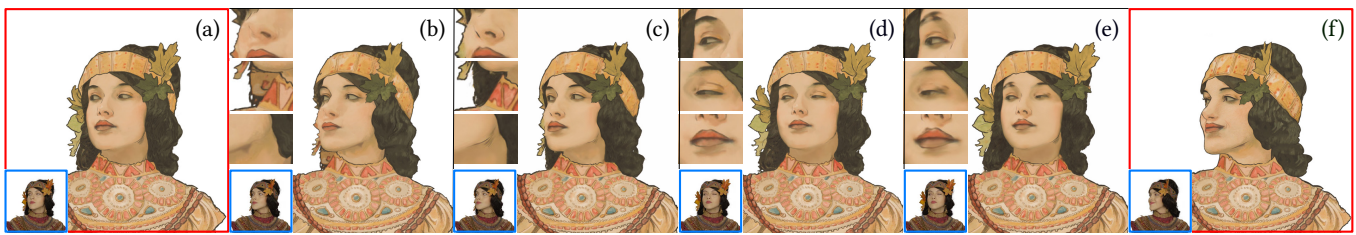}\caption{\revision{Keyframes~(a, f) were
used to stylize the sequence of 154 frames. See the qualitative difference
between Jamri\v{s}ka et al.~\shortcite{Jamriska19}~(b) and our result~(c).
Focusing mainly on zoom-in views, our approach better preserves contour lines
around the nose and chin; moreover, the method of Jamri\v{s}ka et al.~suffers from
blending artifacts---the face is blended into the hair region. On the other
hand, comparison on a different frame from the same sequence shows that the
result of Jamri\v{s}ka et al.~(d) is qualitatively superior to our result~(e)
on this particular frame. See the corresponding zoom-in views where the
approach of Jamri\v{s}ka et al.~produces cleaner results. Video frames~(insets
of a--f) and style exemplars~(a, f) courtesy of \copyright~Muchalogy\permission.}}\label{fig:Res_multi_Imperial}
\end{figure*}

\begin{figure*}[ht]
\def\svgwidth{\hsize}\import{}{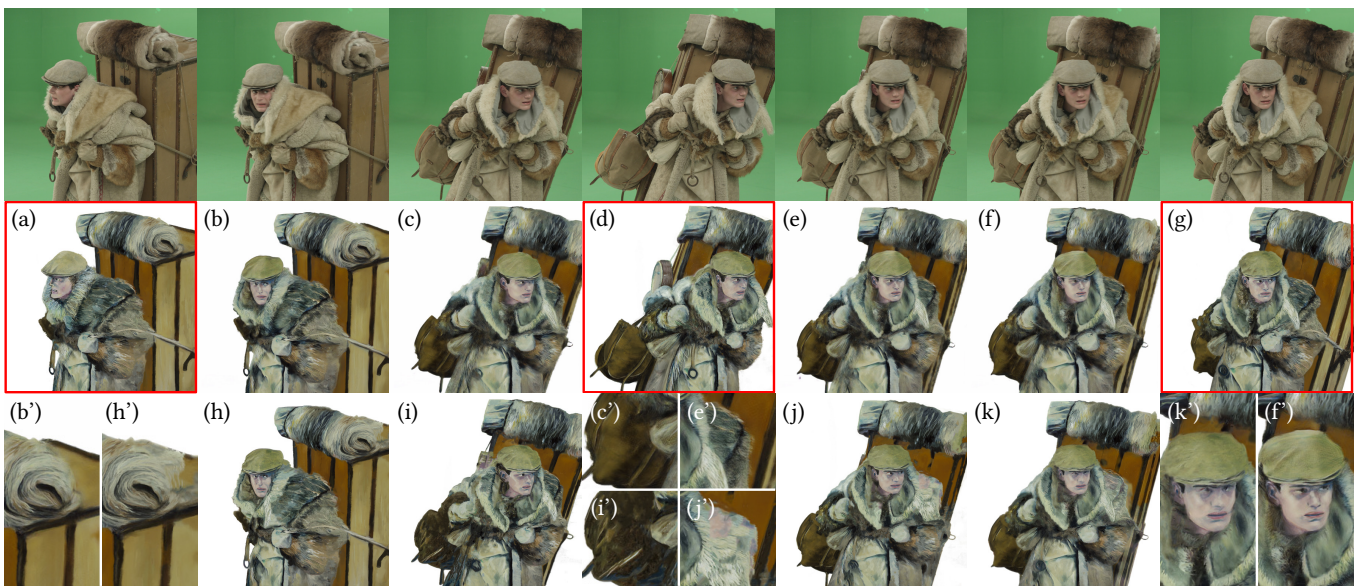}\caption{A complex input sequence (the first
row) with seven keyframes, three of them are shown in~(a, d, g). Here we
compare our approach to the approach of Jamri\v{s}ka et
al.~\shortcite{Jamriska19}. See our result~(b) and theirs~(h) along with the
close-ups~(b', h'); due to their explicit handling of temporal coherence, the
texture of the fur leaks into the box~(h'). Next, compare our result~(c) to
theirs~(i); our approach better reconstructs the bag~(c', i'). Their issue with
texture leakage manifests itself again on the shoulder in~(j, j'), notice how
our approach~(e, e') produces a clean result. Lastly, see how our result~(f, f')
is sharper and the face is better pronounced compared to the result of
Jamri\v{s}ka et al.~\shortcite{Jamriska19}~(k, k'), which suffers from
artifacts caused by their explicit merging of keyframes. Video frames~(top row)
and style exemplars~(a, d, g) courtesy of \copyright~MAUR film\permission.}\label{fig:Res_multi_Krobot}
\end{figure*}

We compared our approach with the current state-of-the-art in keyframe-based
video stylization~\cite{Jamriska19}. For the results see
Figures~\ref{fig:Lim_Zuzka2_head_rotate}, \ref{fig:Res_single_Joli},
\ref{fig:Res_multi_Imperial}, \ref{fig:Res_multi_Krobot}, and our supplementary
video (at 0:08 and 1:08). Note how the overall visual quality, as well as the temporal coherence,
is comparable. In most cases, our approach is better at preserving important
structural details in the target video, whereas the method of Jamri\v{s}ka et
al.~often more faithfully preserves the texture of the original style exemplar.
This is caused by the fact that the method of Jamri\v{s}ka et al.~is
non-parametric, i.e., it can copy larger chunks of the style bitmap to
the target frame. Our method is parametric, and thus it can adapt to fine
structural details in the target frame, which would otherwise be difficult to
reproduce using bitmap chunks from the original style exemplar.

Regarding the temporal consistency, when our full-fledged flicker compensation
based on the mixture of Gaussians is used our approach achieves comparable
coherency in time to the method of Jamri\v{s}ka et al. It is also apparent that
when multiple keyframes are used for stylization, ghosting artifacts mostly
vanish in our method, unlike in Jamri\v{s}ka et al. When the original noisy
sequence is used, or only the bilateral filtering is applied, the resulting
sequence may flicker a little more when compared to the output of Jamri\v{s}ka
et al. However, we argue that the benefits gained from random access and
parallel processing greatly outweigh the slight increase of temporal flicker.
Moreover, the order-independent processing brings also a qualitative
improvement over the method of Jamri\v{s}ka et al.~that tends to accumulate
small errors during the course of the sequence, and visibly deteriorates after
a certain number of frames.

Performance-wise a key benefit of our approach is that once the network is
trained, one can perform stylization of a live video stream in real-time. Even
in the offline setting, when the training phase is taken into account, the
overall end-to-end computation overhead is still competitive. On a 3 GHz
quad-core CPU with Nvidia RTX 2080 GPU, a $512\times 512$ sequence with 100
frames takes around 5 minutes to train until convergence and stylize using our
approach, whereas the method of Jamri\v{s}ka et al.~requires around 15 minutes.

\subsection{Interactive applications}

To evaluate the ideas we presented in practice, we invited artists to work with
our framework. We implement and experiment with three different setups in which
the artists created physical as well as digital drawings. The goal of these
sessions was to stylize one or more video keyframes artistically. Using a
workstation PC, we provided the artists with a version of our framework that
implements real-time interactive stylization of pre-prepared video sequences
and stylization of live camera feeds.

These applications, all of which rely on and strongly benefit from the near
real-time nature of patch-based training as well as the real-time performance
of full-frame inference, naturally lend themselves to fast iteration. The
artist is provided with real-time feedback that approximates what the final
result of video stylization might look like, thus reducing the possibility of
running into issues with artifacts that would be difficult to alleviate later
on.

During the sessions, artists especially appreciated seeing video results very
quickly, as it helps steer creative flow and offers the possibility of
perceiving the effect of individual changes in the style exemplar at a glance.
The overall experience was described as incredibly fun and paradigm-changing,
with little to no negative feedback. Using this system is intuitive and even
suitable for children. These different scenarios are described in detail in the
supplementary material.


\section{Limitations and Future Work}
\label{sec:limits}

Although our framework brings substantial improvements over the
state-of-the-art and makes keyframe video stylization more flexible and
interactive, there are still some limitations that could represent a potential
for further research.

Despite the fact our technique uses different computational machinery than
current state-of-the-art~\cite{Jamriska19} (deep convolutional network
vs.~guided patch-based synthesis), both approaches share similar difficulties
when stylized objects change their appearance substantially over time, e.g.,
when the object rotates and thus reveals some unseen content. Although our
approach often resists slightly longer than patch-based synthesis due to the
ability to generalize better, it usually cannot invent consistent stylization
for new features that were not stylized in the original keyframe,
see~\fg{Lim_Zuzka2_head_rotate}. In this case, the user needs to provide
additional keyframes to make the stylization consistent.

As compared to the method of Jamri\v{s}ka et al.~our approach may encounter
difficulties when processing keyframes at a higher resolution (e.g., 4K) to
stylize high-definition videos. Although the size of patches, as well as the
network capacity, can be increased accordingly, the training may take notably
longer time, as a different multi-scale approach~\cite{Wang18b} could be
necessary. However, the problem of training of larger models is an active
research topic in machine learning, so we believe that soon, more efficient
methods will be developed so that our technique would be applicable also at
higher resolutions.


Although our approach does not require the presence of previous stylized frames
to preserve temporal coherency, the motion-compensated bilateral filter, as
well as the creation of layer with a random mixture of colored Gaussians,
requires fetching multiple video frames. Even though those auxiliary
calculations can still be performed in parallel, they need additional
computation resources. Those may cause difficulties when considering real-time
inference from live video streams. In our prototype, during the live capture
sessions, treatment for improving temporal coherence was not taken into
account. A fruitful avenue for future work would be to implement real-time
variants of the motion-compensated bilateral filter as well as a mixture of
colored Gaussians. Also, different methods could be developed that would enable
the network to keep stylized video temporally coherent without the need to look
into other video frames.

\section{Conclusion}
\label{sec:conclude}
We presented a neural approach to keyframe-based stylization of arbitrary
videos. With our technique, one can stylize the target sequence using only one
or a few hand-drawn keyframes. In contrast to previous neural-based methods,
our method does not require large domain-specific datasets nor lengthy
pre-training. Thanks to our patch-based training scheme, optimized
hyper-parameters, and handling of temporal coherence, a standard appearance
translation network can be trained on a small set of exemplars. Once trained,
it can quickly deliver temporally coherent stylized videos with a visual
quality comparable to the current state-of-the-art in keyframe-based video
stylization, which uses guided patch-based synthesis. A key benefit of our
technique is that it can work in a frame-independent mode, which is highly
beneficial for current professional video editing tools that rely heavily on
random access and parallel processing. It also does not require the explicit
merging of stylized content when slightly inconsistent keyframes are used.

Moreover, since the network in our framework can be trained progressively, and
the inference runs in real-time on off-the-shelf GPUs, we can propose several
new video editing scenarios that were previously difficult to achieve. Those
include stylization of a live video stream using a physical hand-drawn exemplar
being created and captured simultaneously by another video camera. We believe
interactive scenarios such as this will empower the creative potential of
artists and inspire them with new creative ideas.

\ifcamera

\begin{acks}

We thank the reviewers for their insightful comments and suggestions. We are
also grateful to Aneta Texler for her help on manuscript perparation as well as
Zuzana Studen\'{a}, Pavla \& Jolana~S\'{y}korov\'{a}, and Emma Mit\'{e}ran for
participating in the recordings of live sessions. This research was supported
by Snap Inc., the Research Center for Informatics, grant
No.~CZ.02.1.01/0.0/0.0/16\_019/0000765, and by the Grant Agency of the Czech
Technical University in Prague, grant No.~SGS19/179/OHK3/3T/13 (Research of
Modern Computer Graphics Methods).

\end{acks}

\fi

\bibliographystyle{ACM-Reference-Format}
\bibliography{main}
\end{document}